\pdfoutput=1

\documentclass[twocolumn,english]{revtex4}
\usepackage{amssymb,amsmath,graphicx}
\usepackage{color}
\usepackage[T1]{fontenc}
\usepackage{times}
\usepackage{babel}
\begin{document}


\title{
  Adiabatic pumping driven by moving kink and quantum standard ampere 
  in buckled graphene nanoribbon
}

\author{Dominik Suszalski}
\affiliation{Institute for Theoretical Physics,
  Jagiellonian University, \L{}ojasiewicza 11, PL--30348 Krak\'{o}w, Poland}

\author{Adam Rycerz}
\affiliation{Institute for Theoretical Physics,
  Jagiellonian University, \L{}ojasiewicza 11, PL--30348 Krak\'{o}w, Poland}

\date{September 25, 2020}

\begin{abstract}
  A~quantum pump in buckled graphene ribbon with armchair edges is discussed
  numerically. By solving the Su-Schrieffer-Heeger model and performing
  the computer simulation of quantum transport we find that a~kink
  adiabatically moving along the metallic ribbon results in highly-efficient
  pumping, with a charge per kink transition close to the maximal value
  determined by the Fermi velocity in graphene. 
  Remarkably, insulating nanoribbon show the quantized value of a~charge per
  kink ($2e$) in relatively wide range of the system parameters,
  providing a~candidate for the quantum standard ampere.
  We attribute it to the presence of a~localized electronic state, moving
  together with a~kink, whose energy lies within the ribbon energy gap.
\end{abstract}

\maketitle

\section{Introduction}
Nanoscale electromechanical devices based on novel two-dimensional
materials, such as graphene, constitute a~specific class of systems
being interesting both due to their fundamental and technological
aspects \cite{Ben15,Kha17}.
In attempt to improve electromechanical characteristics of such devices,
one need to address fundamental issues concerning the structure of
effective Hamiltonian at nanoscale, including tight-binding parameters
\cite{Kat12}, elastic coefficients \cite{Tsa10,Der18},
and electron phonon-coupling \cite{Voz10}. 
Numerous works have addressed the idea of quantum pumping in graphene
nanostructures \cite{Pra10,Wak10,San12,Jia13,Gri13,Abd17,Fuj20,Zha20,Sus20},
employing various physical mechanisms. These include gate-driven pumping
\cite{Pra10,Wak10}, laser light \cite{San12}, strain-induced fields
\cite{Jia13}, tunable magnetoresistance \cite{Gri13}, quantum Hall states
\cite{Abd17}, but also electromechanical effects accompanying sliding 
Moir\'{e} patterns in twisted bilayer \cite{Fuj20,Zha20}, or 
(most recently) moving kink in buckled graphene nanoribbon \cite{Sus20}. 

As generic quantum pump transfers electric charge between
two reservoirs at zero external bias, solely due to periodic modulation
of the device connecting the reservoirs \cite{Naz09}, new
fundamental and practical aspects of any particular pumping mechanism
may be unveiled with the charge quantization at nanoscale.
Various single-electron pumps were considered as
candidates for quantum standard ampere  \cite{Pek13,Kan16,Poi19}:
In case the charge pumped per cycle is perfectly quantized
(i.e., equal to $Q=ne$, with $n$ integer) in a~considerably wide range of
driving parameters, 
the output current delivered by the device is $I_P=nef_P$, with $f_P$
being the external frequency, and the SI unit of current can be re-defined
by fixing the elementary charge $e$ at
$1.602176634\times{}10^{-19}\,$A$\cdot$s, with the second defined via
the ground-state hyperfine transition frequency of the
cesium 133 atom, $\Delta{}\nu_{\rm Cs}=9\,192\,631\,770\,$Hz \cite{New18,Moh18}. 

So far, single-electron pumps with potential to operate as 
standard ampere are predominantly based on gate-driven quantum dot systems
\cite{Poi19}. We argue here, presenting the results of computer
simulation of quantum transport, that electromechanical pump based on
buckled graphene ribbon, which has recently attracted some attention as
a~physical realization of the classical $\phi^4$ model and its topological
solutions ({\em kinks}) connecting two distinct ground states
\cite{Yam17,Yam19}, may also be considered as a~counterpart to
the above-mentioned single-electron pumps. 

\begin{figure}[!t]
  \includegraphics[width=\linewidth]{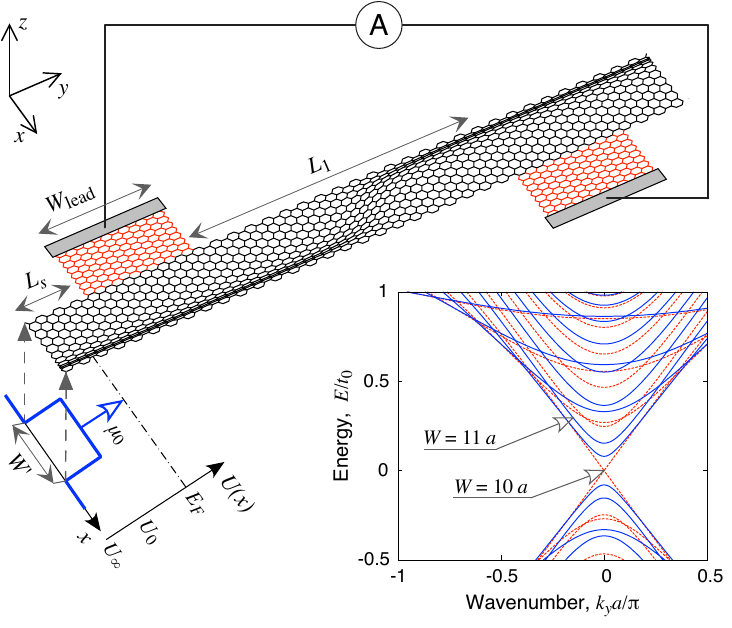}
\caption{ \label{setup2fig}
  Buckled graphene ribbon as a~quantum pump.
  Top: Nanoribbon buckled by changing the distance between fixed armchair
  edges from the equilibrium width of $W=11\,a$
  (with $a=0.246\,$nm the lattice spacing) to $W'=0.9W$, 
  attached to heavily-doped graphene leads (red), each of width $W_\infty$,
  separated by distance $L_1$. The total ribbon length
  is $L=L_1+2W_\infty+2L_s$, where $L_s$ denotes distance between
  the free ribbon edge and the lead edge. The kink is formed near
  the ribbon center. The ammeter detects the current driven by a~moving kink.
  The gate electrode (not shown) is placed underneath to tune the chemical
  potential $\mu_0$ in the ribbon area. The schematic potential profile
  $U(x)$ (bottom left) and the coordinate system (top left) are also shown.
  Inset: Band structure of the infinite flat ribbon with armchair edges
  for $W=11\,a$ (solid blue lines) and  $W=10\,a$ (dashed red lines). 
}
\end{figure}

Earlier \cite{Sus20}, we have shown  that the system similar to
the presented in Fig.\ \ref{setup2fig} consisting of {\em metallic}
graphene nanoribbon with armchair edges coupled to heavily-doped graphene
leads may operate as efficient quantum
pump, but the charge per cycle is not quantized.
Here, the discussion is supplemented by (i) taking the case of 
{\em insulating} nanoribbon into account, and (ii) by optimizing
atomic bond lengths in a~framework of the Su-Schrieffer-Heeger (SSH) model
\cite{Dre98} including electron-phonon coupling of the Peierls type
\cite{Pei55}. 
As a~result, we find that for an insulating ribbon a~single electronic state
localized at the kink is well-separated from extended states, and the charge
pumped is quantized. Topological aspects of the system are crucial
to understand the charge pump operation, since the electron-phonon coupling
leads to the peculiar, arrangement of shorten (lengthen) bonds
being perpendicular (parallel) to the main ribbon axis 
in the kink area, resulting in the electron localization. 

We also show in this paper that, although the kink shape
is well-described within the standard molecular dynamics
potentials for graphite-based systems (as implemented in the LAMMPS package
\cite{Pli95}), for accurate modeling of quantum transport phenomena
one needs to include small corrections to the bond length
(up to a~few percents the kink area), following from electron-phonon
coupling. 
A~minimal quantum-mechanical Hamiltonian, of the SSH type, 
allowing one to model both the kink shape and the transport, is proposed. 

Remaining part of the paper is organized as follows.
In Sec.\ \ref{model}, we present the model Hamiltonian and our method
of approach.
In Sec.\ \ref{qustats}, we discuss quantum states of a~finite section
of buckled ribbon (i.e., {\em closed} system) with a~kink.
The conductance, and the adiabatic pumping in the ribbon coupled to
the leads ({\em open} system) is analyzed numerically
in Sec.\ \ref{condpump}. 
The conclusions are given in Sec.\ \ref{conclu}.

\begin{figure}[!t]
  \includegraphics[width=0.8\linewidth]{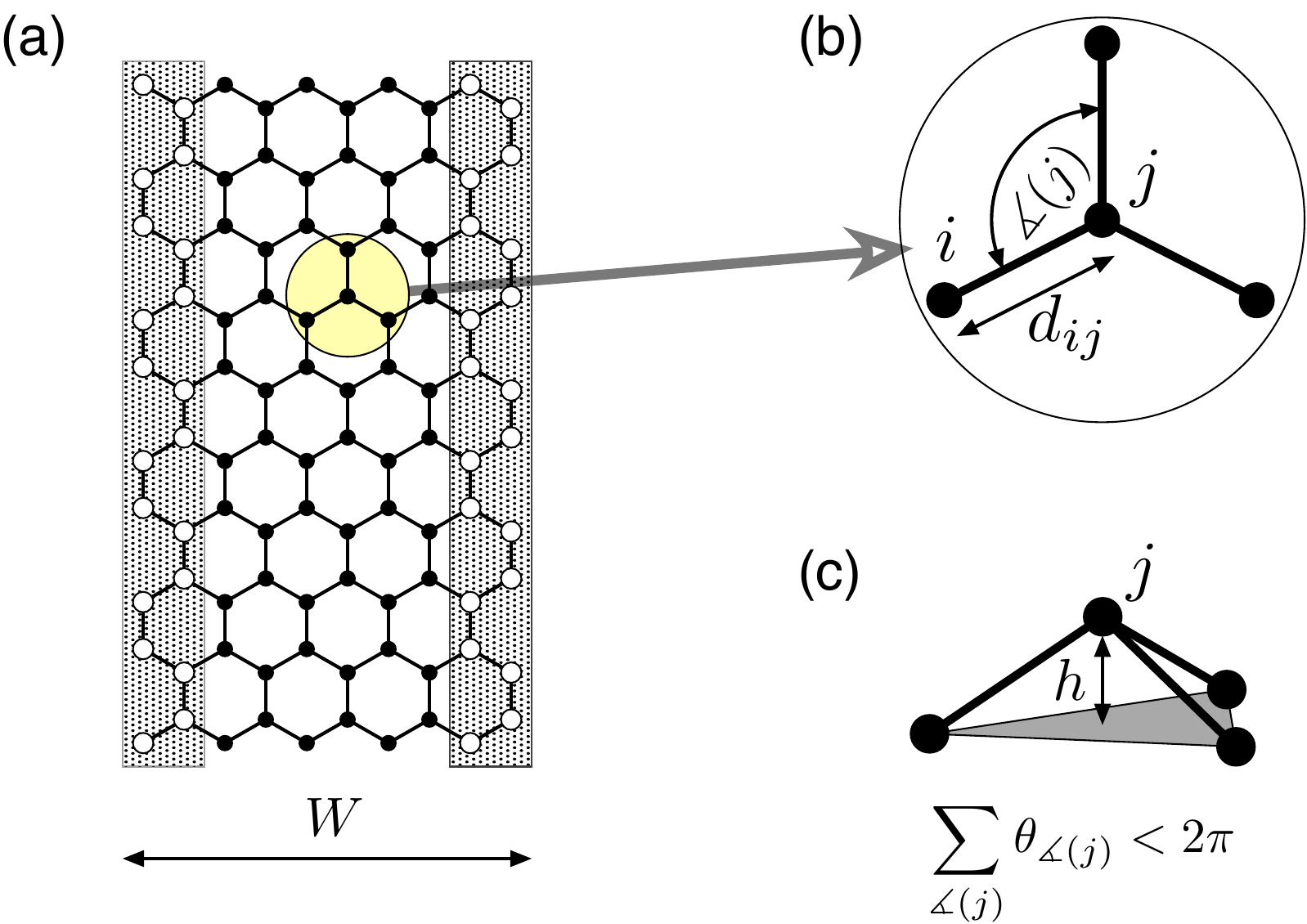}
\caption{ \label{janglefig}
  (a) Short section of a~nanoribbon of $W=5\,a$ width. Two rows of carbon
  atoms near each armchair edge (open symbols on dotted areas) are
  fixed implementing the clamped boundary conditions. Remaning atoms
  (full symbols) are adjusted to minimize the ground-state energy.
  (b) Bond lengths ($d_{ij}$) and angles ($\theta_{\measuredangle(j)}$) used
  in the system Hamiltonian [see Eqs.\ (\ref{hamssh}), (\ref{tkinham}),
  (\ref{vbonds}), (\ref{vangles})].
  (c) The out-of-plane deformation parametrized by the sum of three
  angles with a~common vertex $j$, related to a~tetrahedron height
  $h^2\propto{}2\pi-\sum_{\measuredangle(j)}\theta_{\measuredangle(j)}$. 
}
\end{figure}

\section{Model and methods}
\label{model}

\subsection{The Hamiltonian}
\label{model:ham}

Our analysis starts from the Su-Schrieffer-Heeger (SSH) Hamiltonian
for graphene nanostructures \cite{Dre98,Kar17,Gro18},
with potential energy describing the covalent bonds \cite{Tsa10}
\begin{equation}
  \label{hamssh}
  {\cal H}_{\rm SSH} = T + V_{\rm bonds} + V_{\rm angles},
\end{equation}
where
\begin{align}
  T &=-t_0 \sum_{\langle{}ij\rangle,s}
    e^{-\beta{\delta{}d_{ij}}/{d_0}}
  \left(
    c_{i,s}^\dagger{}c_{j,s}+c_{j,s}^\dagger{}c_{i,s}
  \right),
  \label{tkinham} \\
  V_{\rm bonds} &=
  \frac{1}{2}K_d\sum_{\langle{}ij\rangle}\left(d_{ij}-d_0\right)^2,
  \label{vbonds} \\
  V_{\rm angles} &=
  \frac{1}{2}K_{\theta}\sum_{j} \,
  {\sum_{\measuredangle(j)}}
  \left(\theta_{\measuredangle(j)}-\theta_0\right)^2
  \nonumber\\
  &+ V_\delta \sum_{j}
  \left(2\pi - \sum_{\measuredangle(j)}\theta_{\measuredangle(j)}\right), 
  \label{vangles}
\end{align}
with a~constrain
\begin{equation}
  \label{constdij}
  \sum_{\langle{}ij\rangle}d_{ij}\equiv{}{\cal C}\ \left(=\text{const.}\right). 
\end{equation}

The kinetic-energy operator for $\pi$ electrons $T$ (\ref{tkinham})
includes the hopping-matrix elements ($t_{ij}$) corresponding to the
nearest-neighbors on a~honeycomb lattice (denoted by using brackets
$\langle{ij}\rangle$), with the equilibrium hopping
integral $t_0=2.7\,$eV.
The change in bond length, $\delta{}d_{ij}=d_{ij}-d_0$, is calculated
with respect to the equilibrium bond length $d_0=a/\sqrt{3}$, with
$a=2.46\,$\AA$\ $ being the lattice spacing. 
The operator $c_{i,s}^\dagger{}$ (or $c_{i,s}$) creates (or annihilates)
a~$\pi$ electron at the $i$-th lattice site with spin $s$.
The electron-phonon coupling, quantified by the dimensionless parameter
$\beta=-\left.\partial{}\ln{}t_{ij}/\partial\ln{}d_{ij}\right|_{d_{ij}=d_0}$ 
(to be specified later),
is represented by the exponential factor in Eq.\ (\ref{tkinham}) replacing
standard Peierls form $(1-\beta\delta{}d_{ij}/d_0)$ in order to prevent
$t_{ij}$ from changing the sign upon strong lattice deformation. 

The next two terms in Eq.\ (\ref{hamssh}), $V_{\rm bond}$ (\ref{vbonds})
and $V_{\rm angles}$ (\ref{vangles}), approximates the potential
energies for the bond stretching and bond angle bending (respectively); 
see Fig.\ \ref{janglefig}.
The parameters $K_d=40.67\,$eV$/$\AA$^2$, $K_\theta=5.46\,$eV$/$rad$^2$,
and $\theta_0=\pi/3$, are taken from Ref.\ \cite{Tsa10} and restore
the actual in-plane elastic coefficients of bulk graphene in the case of 
$\beta=0$. Otherwise (for $\beta\neq{}0$), a~correction to the potential
energy per bond can be estimated as 
\begin{align}
  \left({N_{\rm b}}\right)^{-1}&\left.
  \frac{\partial^2{\langle{}T\rangle}}{\partial{}d_{ij}^2}
  \right|_{\{d_{ij}=d_0\}} = \frac{\beta{}t_0}{d_0^2}\sum_s\left\langle
  c_{i,s}^\dagger{}c_{j,s}+c_{j,s}^\dagger{}c_{i,s}
  \right\rangle \nonumber\\
  \approx{}& \beta\times{}1.405\mbox{ eV/{\normalfont\AA}$^2$}
  \ \ \ll{}\ \ K_d\ \ \ \ (\text{for }\ \ \beta\sim{}1),  
\end{align}
with the number of C-C bonds $N_b$. 
The second approximate equality in the above is obtained by
substituting
$\sum_s\left\langle{}c_{i,s}^\dagger{}c_{j,s}+c_{j,s}^\dagger{}c_{i,s} %
\right\rangle\approx{}1.050$
(with $i$ and $j$ the nearest neighbors),
being the value for a~perfect, bulk graphene
sheet at the half electronic filling \cite{Mar97foo}. 

The expression for $V_{\rm angles}$ (\ref{vangles}) consists of two terms,
each involving summation over the three angles $\measuredangle(j)$ having
a~common vertex at a~given lattice site $j$ (see Fig.\ \ref{janglefig}). 
First term, $\propto{}K_\theta$, represents the harmonic approximation for
in-plane bond angle bending. For out-of plane deformations, quantified
by the height $h_j$ of a~tetrahedron formed by $j$-th site and its three
nearest neighbors, this term represents a~fourth-order correction to the
potential energy.
A~realistic description of out-of-plane deformations requires a~correction
of the $\sim{}h_j^2$ order (for $h_j\ll{}d_0$).
Here we propose a~term proportional to the excess angle,
\begin{equation}
  \delta_j = 2\pi - \sum_{\measuredangle(j)}\theta_{\measuredangle(j)}
  \approx 3\sqrt{3}\left(\frac{h_j}{d_0}\right)^2,
\end{equation}
with the coefficient $V_\delta\approx{}t_0=2.7\,$eV roughly
approximating the bending rigidity of graphene
\cite{Der18,Voz10}.  
The main advantage of such an approach is that it requires no computationally
expensive operations since the four-body term ($\propto{}V_\delta$) depends
only on angles ($\theta_{\measuredangle(j)}$) earlier determined for the three-body
term ($\propto{}K_\theta$).
The validity of our approach, in comparison with standard molecular dynamics
treatments \cite{Yam17,Yam19}, is discussed later in this paper.

\subsection{The optimization procedure}
\label{model:opt}

Throughout the paper, we compare the results obtained in the absence of
electron phonon coupling, $\beta=0$, and for the dimensionless parameter
$\beta=3$; the two values chosen to bound the possible range of $\beta$
(see Ref.\ \cite{Voz10}).
It is worth to stress here that in the fortcoming analysis
physical properties are discused as functions of the deformation applied,
and thus the other choice of $\beta\neq{}0$ (being the propotionality
coefficient between the local deformation and corresponding correction
to the Hamiltonian) may rather shift the characteristic features observed
than change the picture in a~qualitative manner. 

In the $\beta=0$, electronic and lattice degrees of freedom
are decoupled, and one simply need to solve a~purely classical minimization
problem for the potential energy part of the Hamiltonian ${\cal H}_{\rm SSH}$
(\ref{hamssh}), given by $V_{\rm bonds}+V_{\rm angles}$ [see Eqs.\ (\ref{vbonds})
and (\ref{vangles})]. 
For the $\beta\neq{}0$ case, the average kinetic energy can be calculated as
\begin{align}
  \langle{T}\rangle &=
  -t_0 \sum_{\langle{}ij\rangle,s}\exp\left(
    -\beta\frac{\delta{}d_{ij}}{d_0}
  \right)\left\langle
    c_{i,s}^\dagger{}c_{j,s}+c_{j,s}^\dagger{}c_{i,s}
  \right\rangle \nonumber \\
  &= 2\sum_{ij}t_{ij}\sum_{1\leqslant{}k\leqslant{}N_{\rm el}/2}
  \left[\psi_k^{(i)}\right]^{\star}\psi_k^{(j)} \nonumber \\
  &= 2\sum_{1\leqslant{}k\leqslant{}N_{\rm el}/2}E_k, \label{ttaver}
\end{align}
where the factor $2$ in the last two expressions follows from a~spin
degeneracy, $t_{ij}=-t_0\exp(-\beta{}\delta{}d_{ij}/d_0)$ if $i$ and
$j$ are the nearest neighbors (otherwise, $t_{ij}=0$), and
$\psi_k^{(j)}$ denotes the probability amplitude for the $k$-th
eigenstate of the kinetic energy operator $T$ (\ref{tkinham}) at
$j$-th lattice site. We further suppose that the eigenstates are ordered
such that the energies $E_1\leqslant{}E_2\leqslant{}\dots\leqslant{}
E_{N_{\rm at}}$, with the number of atoms $N_{\rm at}$, and that the number of
electrons $N_{\rm el}$ is even for simplicity. 

In both cases ($\beta=0$ and $\beta=3$), the numerical minimization of
the ground-state energy
\begin{equation}
\label{egdef}
E_G=E_G(\{{\bf R}_j\})=\langle{\cal H}_{\rm SSH}\rangle,
\end{equation}
with respect to atomic positions $\{{\bf R}_{j}\}$, is performed employing
the modified periodic boundary conditions in $y$-direction (see Fig.\
\ref{setup2fig}).  Namely, the system is invariant upon $y\mapsto{}y+L$
and $z\mapsto{}-z$, forcing the kink formation in a~buckled ribbon.
The outermost two rows of atoms near each armchair edge are fixed
during the minimization (see Fig.\ \ref{janglefig}), 
and buckling of the ribbon is realized by changing the distance between
the fixed edges [see Fig.\ \ref{janglefig}(a)]  from $W$ to $W'<W$. 
Furthermore, the number of electrons is fixed at $N_{\rm el}=N_{\rm at}$,
with $N_{\rm at}=3600$
for metallic armchair ribbon ($W=10\,a$, $L=90\sqrt{3}\,a$) or
$N_{\rm at}=3960$ for insulating armchair ribbon ($W=11\,a$, $L=90\sqrt{3}\,a$).
Although in open system, coupled to the leads, the average
$N_{\rm el}$ varies with the chemical potential, such fluctuations
(typically, limited to $\Delta\mu<0.1\,$eV or, equivalently,
$\Delta{}N_{\rm el}/N_{\rm at}\approx{}0.18\,(\Delta{}\mu/t_0)^2<2\times{}10^{-4}$;
see Ref.\ \cite{Ryc16}) are insignificant when determining the optimal bond
lengths. 
Alternatively, one can interpret the $\beta=0$ case as a~hypothetical
$N_{\rm el}=0$ situation, leading to bond length modifications not exceeding
a~few percents (see below). 

For a~fixed $W'/W$ ratio and the kink position $y=y_0$, the computations
proceed as follows. 

The initial arrangement of carbon atoms is given by
\begin{equation}
\label{rr0js}
   {\bf R}_j^{(0)}=\left(x_j^{(0)},y_j^{(0)},z_j^{(0)}\right),
   \ \ \ \ \ \ 
   j=1,2,\dots,N_{\rm at}, 
\end{equation}
where
\begin{align}
  x_j^{(0)} &=  X_{W,W'}(\tilde{x}_j),
  \nonumber \\
  y_j^{(0)} &= \tilde{y}_j,
  \label{xyz0js} \\
  z_j^{(0)} &= H\tanh\left(\frac{\tilde{y}_j-y_0}{\Lambda}\right)
  \sin^2\left(\frac{\pi\tilde{x}_j}{W}\right),
  \nonumber 
\end{align}
with $(\tilde{x}_j,\tilde{y}_j)$ being the coordinates of $j$-th atom on
a~flat honeycomb lattice, and the scaling function
\begin{equation}
X_{W,W'}(x) = \begin{cases}
  x, & \text{for }\ \ x<0, \\
  (W'/W)\,x, & \text{for }\ \ 0\leqslant{}x<W, \\
  x\!-\!W\!+\!W', & \text{for }\ \ x\geqslant{}W. 
\end{cases}
\end{equation}
The buckle height $H$ in Eq.\ (\ref{xyz0js}) is adjusted such that
${\cal C}=N_{\rm b}d_0$ in Eq.\ (\ref{constdij}).
The kink size is fixed at $\Lambda=5\,a$, roughly approximating the kink
profiles reported in Refs.\ \cite{Yam17,Yam19}. 

At first step, we minimize the potential energy term  
$V_{\rm bonds}+V_{\rm angles}$, ignoring a~constrain given by Eq.\ (\ref{constdij}).
This gives us the solution for $\beta=0$ in the Hamiltonian
${\cal H}_{\rm SSH}$ (\ref{hamssh}).

Next step, performed only if $\beta>0$, involves a~further adjustment of atomic
positions $\{{\bf R}_j\}$ such that full ground-state energy $E_G$
(\ref{egdef}) reaches a~minimum. In practice, we determine hopping parameters
$\{t_{ij}\}$, wavefunctions $\left\{\psi_k^{(j)}\right\}$, and correlation
functions
$\left\{\langle{}c_{is}^{\dagger}c_{js}\rangle\right\}$ [see Eq.\ (\ref{ttaver})]
for given $\{{\bf R}_j\}$-s, and then find (within the gradient descent
method) a~conditional minimum of $E_G$ with respect to
$\{{\bf R}_j\}$ at fixed values of $\left\{\langle{}c_{is}^{\dagger}c_{js}
\rangle\right\}$-s, 
satisfying a~constrain given by Eq.\ (\ref{constdij}).
The procedure is iterated until the numerical convergence is reached. 
Typically, after $3$--$4$ iterations the atomic positions $\{{\bf R}_j\}$-s
are determined with the accuracy better then $10^{-5}\,a$.

\subsection{Comparison with LAMMPS results}

\begin{figure}[!t]
\centerline{\includegraphics[width=\linewidth]{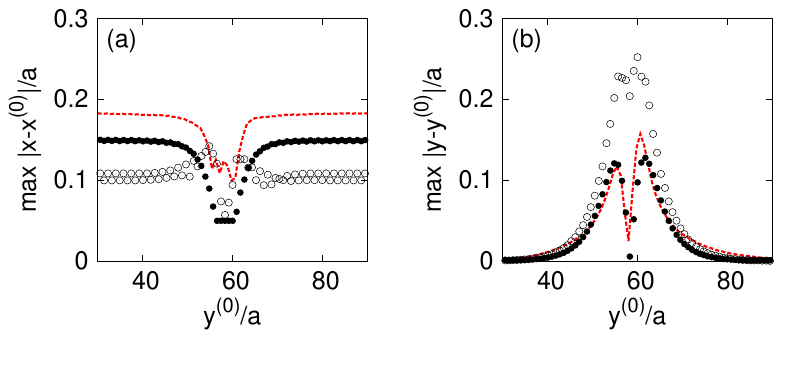}}

\centerline{\includegraphics[width=\linewidth]{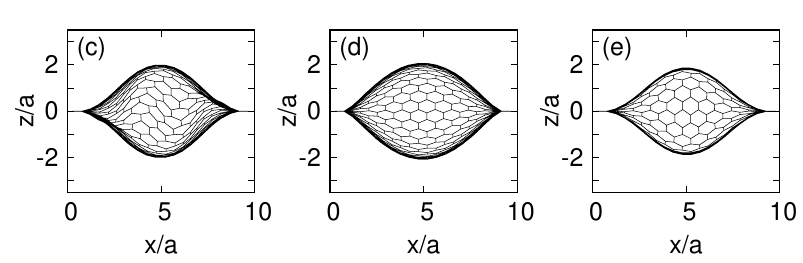}}
\caption{ \label{lammpsfig}
  Spatial arrangement of carbon atoms after the optimization procedure
  presented in Sec.\ \ref{model:opt} compared with the LAMMPS output. 
  (a,b) Maximal displacement of an atom along (a) $x$-axis or
  (b) $y$-axis in the coordinate system of Fig.\ \ref{setup2fig},
  after the optimization starting from  initial coordinates
  $x=x^{(0)}$, $y=y^{(0)}$ [see Eqs.\ (\ref{rr0js}), (\ref{xyz0js})].
  Open (or closed) symbols correspond to $\beta=0$ (or $\beta=3$).
  Red dashed lines show the LAMMPS results for comparison \cite{Pli95,lamm2}.
  (c)--(e) C-C bonds after the optimization projected onto the $x$--$z$ plane
  for (c) $\beta=0$, (d) $\beta=3$, and (e) the LAMMPS results.
  [Notice that difference in scales for $x$ and $z$ axes  
  visually amplifies the buckling.] 
  The system parameters are $L=90\sqrt{3}\,a$, $W=11\,a$, $W'/W=0.9$, and
  the kink position is $y_0=(3/8)L\approx{}58.5\,a$ for all cases. 
}
\end{figure}

A~brief comparison of the kink shape following from the numerical procedure 
described above with the corresponding output produced by the LAMMPS
Molecular Dynamics Simulator \cite{Pli95,lamm2} is presented in
Fig.\ \ref{lammpsfig}.

In order to quantify the difference in atomic arrangements obtained within
different approaches, we choose the maximal absolute displacement of atom
along the $x$ (and $y$) axis, max$\,|x-x^{(0)}|$ (and max$\,|y-y^{(0)}|$),
where the maximum is taken for a~subset of atoms with equal initial
$y^{(0)}$ coordinates; see Figs.\ \ref{lammpsfig}(a) and \ref{lammpsfig}(b). 
It is sufficient to display the data corresponding to a~vicinity of the kink,
$30\leqslant{}y^{(0)}/a\leqslant{}90$, since far away from the kink position
(being fixed at $y_0\approx{}58.5\,a$) both the quantities considered become
$y^{(0)}$-independent. (As free boundary conditions are applied in case
the LAMMPS package is utilized, some $y^{(0)}$-dependences reappear 
near the free zigzag edges, but they are much smaller in magnitude 
than dependencies in the kink area.)

It is clear from Figs.\ \ref{lammpsfig}(a) and \ref{lammpsfig}(b) that the
LAMMPS results (see red-dashed lines) are closer to the obtained
with our optimization procedure in the presence of electron-phonon coupling,
$\beta=3$ (solid symbols), then for $\beta=0$ (open symbols).
Also, $x$--$z$ views of the system, presented in Figs.\ \ref{lammpsfig}(c),
\ref{lammpsfig}(d), and \ref{lammpsfig}(e), show that approximate mirror
symmetry of the kink appears for $\beta=3$ [see Fig.\ \ref{lammpsfig}(d)]
and for the LAMMPS results [Fig.\ \ref{lammpsfig}(e)], but is absent
for $\beta=0$ [Fig.\ \ref{lammpsfig}(c)]. 

The above observations can be rationalized taking into accout that
four-body (dihedral) and long-range Lennard-Jones potential energy terms
are included in the LAMMPS package but absent in our model Hamiltonian
${\cal H}_{\rm SSH}$ (\ref{hamssh}). In the presence of electron-phonon
coupling ($\beta>0$), however, the average kinetic energy
$\langle{}T\rangle$ (\ref{ttaver}) can be interpreted as an effective
long-range (and ``infinite-body'') attractive interaction between atoms,
restoring some features related to the Lennard-Jones forces in molecular
dynamics (including an approximate mirror symmetry of the kink).

\section{Quantum states in closed system with periodic boundary conditions}
\label{qustats}

\subsection{Bond-length modulation}

\begin{figure}[!t]
  \centerline{\includegraphics[width=0.9\linewidth]{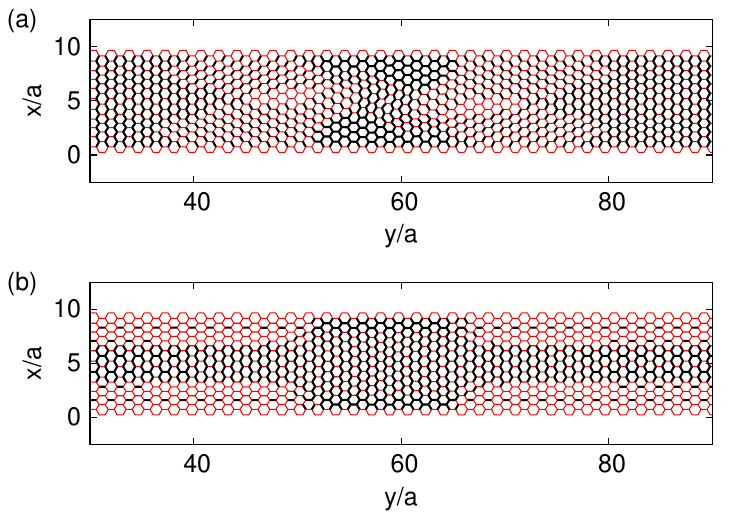}}
  \medskip
  
  \centerline{\includegraphics[width=\linewidth]{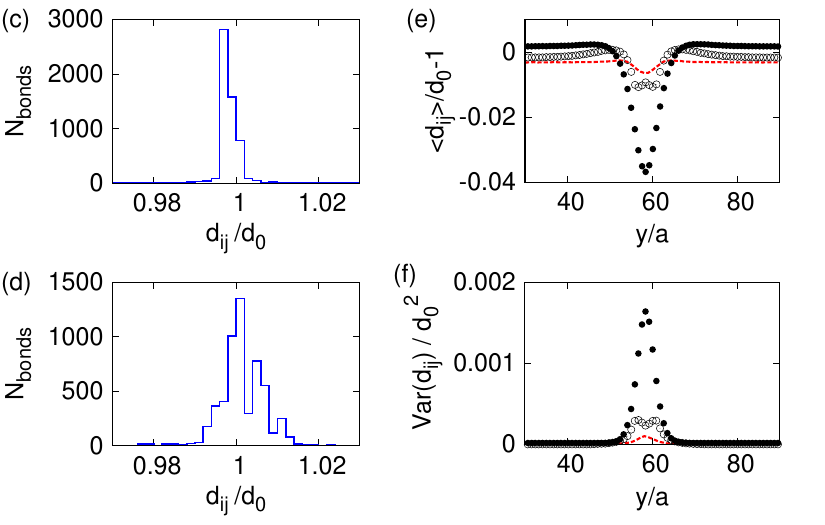}}
  
\caption{ \label{bondfig}
  Bond-length modulation for the same system parameters as in
  Fig.\ \ref{lammpsfig}.
  (a) Shorten (thick black) and lengthen (thin red) bonds projected on
  the $(x,y)$ plane for in the absence of electron-phonon coupling 
  [i.e., $\beta=0$ in Eq.\ (\ref{hamssh})].  
  (b) Same as (a) but for $\beta=3$.  
  (c) and (d) Statistical distributions of the relative bond length
  for $\beta=0$  and $\beta=3$.
  (e) Average bond length distortion displayed as a~function of $y$
  coordinate for $\beta=0$ (open circles) and $\beta=3$ (full circles).
  Red dashed line shows the LAMMPS results. 
  (f) Bond length variation vs.\ $y$ presented with the same symbols (lines)
  as in (e). 
}
\end{figure}

Before discussing the electronic structure of the system, we briefly 
describe small corrections to the bond lengths appearing 
in the kink area due to electron-phonon coupling (see Fig.\ \ref{bondfig}),
which are essential to understand the results presented in the remaining
parts of the paper.

In Figs.\ \ref{bondfig}(a) and \ref{bondfig}(b) we visualize the spatial
arrangements of shorten and lengthen bonds; namely, 
$d_{ij}<\langle{}d_{ij}\rangle{}_{i,j=1\dots{}N_{\rm at}}$ (thick black lines)
and $d_{ij}>\langle{}d_{ij}\rangle{}$ (thin red lines),  
where the average bond length $\langle{}d_{ij}\rangle{}=0.998\,d_0$ for
$\beta=0$ and $W'/W=0.9$, or $\langle{}d_{ij}\rangle{}\equiv{}d_0$ for
$\beta=3$ due to a~constrain imposed [see Eq.\ (\ref{constdij})].
Apparently, in the presence of electron-phonon coupling ($\beta=3$)
a~large rectangular block is formed in the kink area [i.e., for
$|y-y_0|\lesssim{}7\,a$; see Fig.\ \ref{bondfig}(b)],
in which almost
all bonds oriented in the zigzag direction are shorten (resulting in the
hopping element $|t_{ij}|>t_0$) and almost all remaining bonds are 
lengthen ($|t_{ij}|<t_0$). In the absence of electron-phonon coupling
($\beta=0$) the situation is less clear [Fig.\ \ref{bondfig}(a)],
with a~few smaller blocks
of shorten or lengthen bonds forming more complex patterns, some of which
are isotropic, and some show  various crystallographic orientations. 

The qualitative finding presented above is further supported with 
statistical distributions of the relative bond length ($d_{ij}/d_0$),
determined using all $N_{\rm b}=5749$ bonds in the system (for $L=90\sqrt{3}\,a$
and $W=11\,a$) and displayed in Figs.\ \ref{bondfig}(c) and \ref{bondfig}(d). 
In particular, the distribution for $\beta=3$ [Fig.\ \ref{bondfig}(d)]
is significantly wider than for $\beta=0$ [Fig.\ \ref{bondfig}(c)].
Also, bimodal structure of the distribution is visible
in the presence of electron-phonon coupling, suggesting that two distinct
populations of shorten and lengthen bonds are formed in this case. 

The position dependence of the bond-length modulation is illustrated
in Figs.\ \ref{bondfig}(e) and \ref{bondfig}(f), where we display
the mean $\langle{}d_{ij}\rangle{}$ the variance Var$\,(d_{ij})$
calculated for bonds connecting atoms in a~single zigzag line (parallel
to the $x$ direction) as functions
of a~mean $y$-position of carbon atoms in the line.
It is clear that significant bond-length modulations (however, not
exceeding a~few percents of $d_0$) appear only in a~small
vicinity of the kink position, $|y-y_0|\lesssim{}7\,a$, and that the
modulations and noticeably stronger for $\beta=3$ (solid symbols) than
for $\beta=0$ (open symbols).
Remarkably, the LAMMPS results (red-dashed lines) now indicate much weaker
bond-length modulations than our numerical results (regardless $\beta=0$
or $\beta=3$) but are significantly closer to the $\beta=0$ results then
to the $\beta=3$ results.
This shows that the quantum-mechanical nature of the Hamiltonian
${\cal H}_{\rm SSH}$ (\ref{hamssh}), relevant for spacial arrangement
of carbon atoms only if $\beta\neq{}0$,
is crucial for an accurate description of bond-lengths corrections.

\begin{figure}[!t]
  \includegraphics[width=0.9\linewidth]{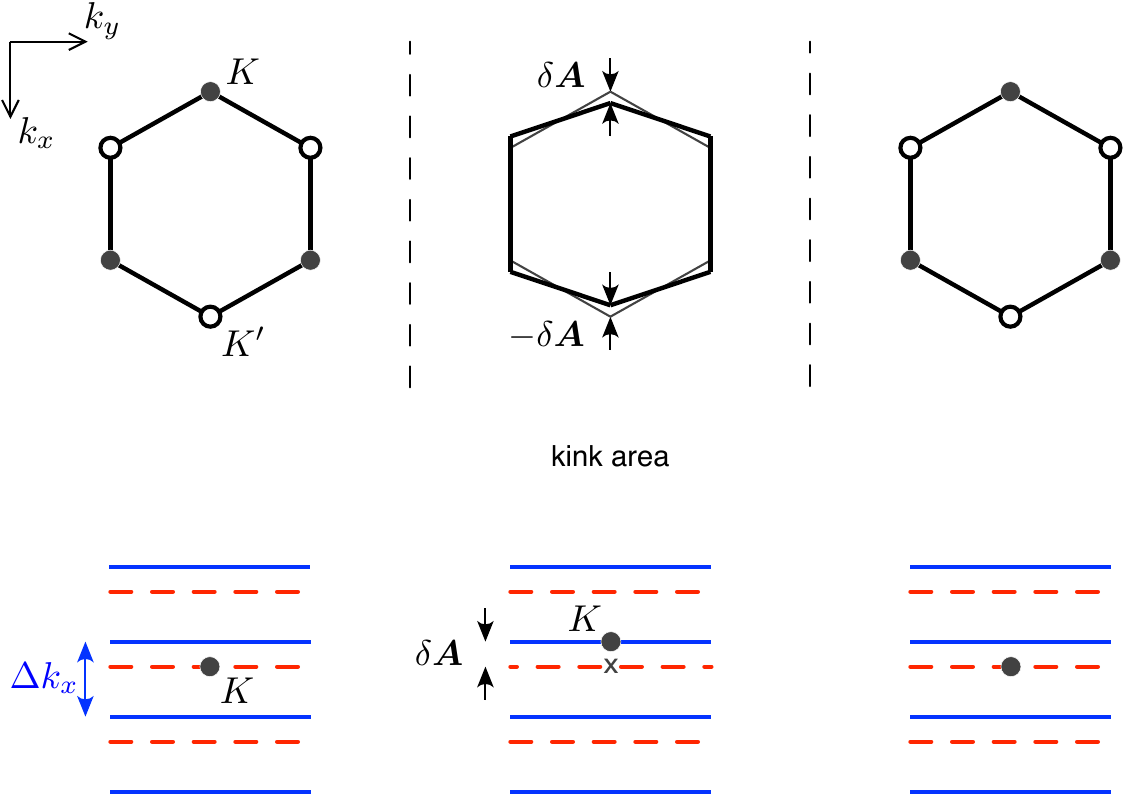}
\caption{ \label{fbzfig}
  Schematic illustration of the mechanism of current blocking
  (or electron localization) in metallic (or insulating) graphene
  nanoribbon with a~kink.
  Top: 
  Relative modification of the bond lengths in the kink area shifts 
  each Dirac point ($K$ and $K'$) in the momentum space
  by $\pm\delta\mbox{\boldmath$A$}$.
  Bottom: Transverse momentum quantization near the $K$ point (blue solid
  or red dashed lines, separated by $\Delta{}k_x$, correspond to the insulating
  or the metallic case, respectively) combined with the $K$-point shift may
  locally turn metallic nanoribbon into insulating one or {\em vice versa}. 
}
\end{figure}

\subsection{The current blocking}

In order to understand how the bond-length modulation may affect the
transport properties, we focus now on the Dirac points ($K$ and $K'$)
and changes in their positions in the first Brillouin zone 
due to strain-induced fields (see Fig.\ \ref{fbzfig}). 

Revisiting the derivation of an effective Dirac equation for graphene
one finds that weak deformations introduce peculiar gauge fields,
with the vector potential for $K$ valley \cite{Voz10}
\begin{equation}
\label{akstrain}
  {\mbox{\boldmath ${A}$}}_K \equiv
  \left(\begin{array}{c}
    {A}_{K,x}  \\  {A}_{K,y} 
  \end{array}\right)
  = 
  \frac{c\beta}{d_0}
  \left(\begin{array}{c}
    u_{xx}-u_{yy}  \\  -2u_{xy} 
  \end{array}\right), 
\end{equation}
where $c$ is a dimensionless coefficient of the order of unity,
$u_{ij}=\frac{1}{2}(\partial_i{}u_j+\partial_j{}u_i)$ (with $i,j=x,y$)
is the symmetrized strain tensor for in-plane deformations \cite{Lan59}
and the coordinate system is chosen as in Fig.\ \ref{setup2fig} (i.e.,
such that the $x$ axis corresponds to a zigzag direction of a honeycomb
lattice). For the $K'$ valley, the strain-induced field has an opposite sign 
(namely, ${\mbox{\boldmath ${A}$}}_{K'}=-{\mbox{\boldmath ${A}$}}_K$). 

For an approximately uniform compression along the $x$ direction occurring
in the kink area, we have $u_x\approx{}(W'/W)x$, $u_y=y$, and the $K$
point is shift by 
$\delta{}\mbox{\boldmath ${A}$}\propto{}-(1-W'/W)\hat{k}_x$ 
with $\hat{k}_x$ being a~unit vector in the $k_x$ direction, while
the $K'$ point is shift by $-\delta\mbox{\boldmath ${A}$}$,
as visualized in top panels of Fig.\ \ref{fbzfig}. 
Away from the kink area, buckling without changing bond lengths
does not create strain-induced fields 
($\delta{\mbox{\boldmath ${A}$}}\approx{}0$) \cite{scalarfoo}.  

Additionally, a~finite size along the $x$ direction introduces the well-known
geometric quantization, with the discrete values of quasimomentum $k_x$,
separated by $\Delta{}k_x\sim{}\pi/W$ (see bottom panels in
Fig.\ \ref{fbzfig}).
In principle, for a~particular combination of $\delta\mbox{\boldmath ${A}$}$
and $\Delta{}k_x$, a~nanoribbon may locally change its character from
metallic to insulating (or {\em vice versa}).
In more general situation, if $\delta\mbox{\boldmath ${A}$}$ and
$\Delta{}k_x$ are not precisely adjusted to alter the system properties at
$E=0$, one can find some finite energies ($E>0$ for electrons or $E<0$ for
holes), for which quantum states are available only away from the kink
area (or only in the kink area).
A~direct illustration is provided with the density of states discussed next.

\subsection{Density of states}

\begin{figure}[!t]
  \includegraphics[width=\linewidth]{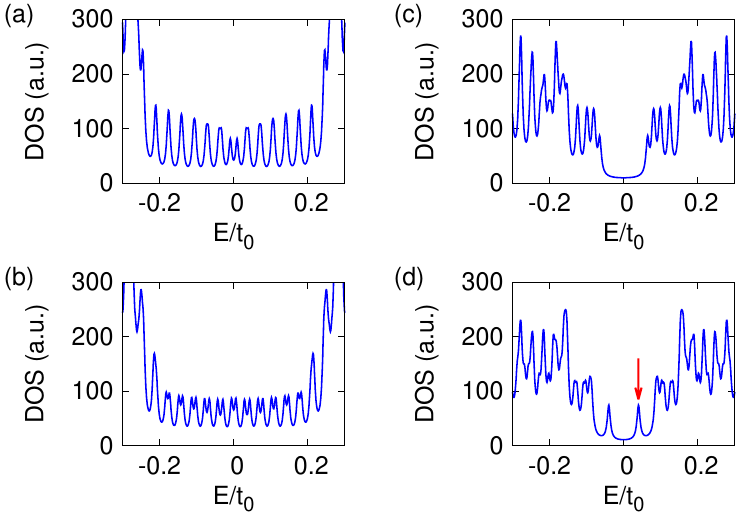}
\caption{ \label{dos4fig}
  Electronic density of states as a~function of energy for a~finite section
  ($L=90\sqrt{3}\,a$) of (a,b) metallic [$W=10\,a$] and (c,d) insulating
  [$W=11\,a$] ribbons with armchair edges, $W'/W=0.9$, and a~single kink.
  The bond lengths are optimized for (a,c) $\beta=0$ or (b,d) $\beta=3$.
  Arrow in (d) indicates one of the localized states, at $E=0.04\,t_0$,
  appearing in the gap area.
  The level broadening parameter [see Eq.\ (\ref{delrep})] is
  $\epsilon=5\cdot{}10^{-3}\,t_0$. 
}
\end{figure}

\begin{figure}[!t]
  \includegraphics[width=\linewidth]{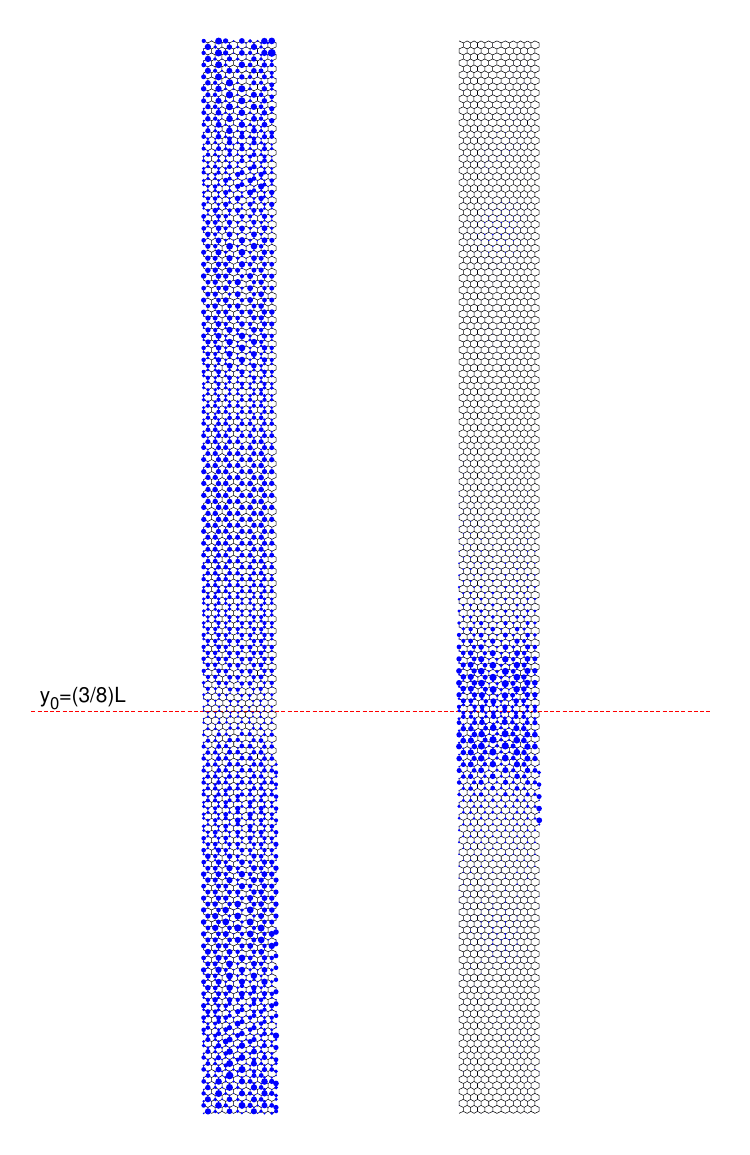}
\caption{ \label{ldosfig}
  Local density of states for $E=0.04\,t_0$ for $W=10\,a$ (left)
  and $W=11\,a$ (right).
  Red horizontal line indicates the kink position.
  The bond lengths are optimized for $\beta=3$. 
  Remaining system parameters are same as in Fig.\ \ref{dos4fig}. 
}
\end{figure}

We consider here two nanoribbons with armchair edges, one of the width
$W=10\,a$ (the {\em metallic} case) and the other of $W=11\,a$
(the {\em insulating} case). The system length is $L=90\sqrt{3}\,a$
in both cases, with modified periodic boundary conditions (see
Sec.\ \ref{model:opt}) applied for both lattice and electronic degrees
of freedom. The two values of $\beta=0$ and $\beta=3$ in the Hamiltonian
${\cal H}_{\rm SSH}$ (\ref{hamssh}) are considered;
the buckling magnitude is fixed at $W'/W=0.9$.
The above parameters allow us to define the two energy scales:
The subband splitting
\begin{equation}
\label{delenw}
  \Delta{}E_{W}=\hbar{}v_F\Delta{}k_x\approx{}0.3\,t_0, 
\end{equation}
and the longitudinal quantization 
\begin{equation}
  \Delta{}E_{L}=2\pi{}\hbar{}v_F/L\approx{}0.03\,t_0. 
\end{equation} 

In Fig.\ \ref{dos4fig}, we display the electronic density of states 
\begin{equation}
\label{dosdef}
  \rho(E) = \sum_n \delta(E-E_n), 
\end{equation}
with $E_n$ denoting the $n$-th eigenvalue of the kinetic-energy operator
$T$ given by Eq.\ (\ref{tkinham}), for all four combinations of $\beta$ and $W$.
For plotting purposes, the $\delta$ function is smeared by a~finite
$\epsilon$; namely, we put 
\begin{equation}
  \label{delrep}
  \delta(x) \rightarrow{} \frac{1}{\pi}\frac{\epsilon}{x^2+\epsilon^2},  
\end{equation}
where $\epsilon=5\cdot{}10^{-3}\,t_0$. 

Since $\Delta{}E_W\gg{}\Delta{}E_L$, metallic [see Figs.\ \ref{dos4fig}(a)
and \ref{dos4fig}(b)] or insulating [see Figs.\ \ref{dos4fig}(c)
and \ref{dos4fig}(d)] character of the ribbon can still be recognized from
the $\rho(E)$ spectrum of its finite section: in the former case, $\rho(E)$
is elevated for any $E$, whereas in the later case, we have
$\rho(E)\approx{}0$ in a~vicinity of $E=0$.

The effects of electron-phonon coupling can be summarized as follows.
In the metallic case, bond length modulation results in small splittings of
the electronic levels [see Fig.\ \ref{dos4fig}(b) for $\beta=3$],
originally showing approximate degeneracy
[see Fig.\ \ref{dos4fig}(a) for $\beta=0$], 
due to amplified scattering between the $k_y$ and $-k_y$ states occurring
in the kink area. 
In the insulating case, we have two energy levels, appearing for
$\beta=3$ [see Fig.\ \ref{dos4fig}(d)] but absent for $\beta=0$
[see Fig.\ \ref{dos4fig}(c)], one for electrons (marked with red arrow)
and one for holes, which occur in the gap range and are well-separated
from other levels, suggesting that they are associated with localized states. 

The above expectation is further supported with local density of states 
(presented Fig.\ \ref{ldosfig}) 
\begin{equation}
\label{locdosdef}
  \rho_{\rm loc}({\bf R}_j,E) =
  \sum_n \left|\psi_n^{(j)}\right|^2 \delta(E-E_n), 
\end{equation}
where the $\delta$ function is represented via Eq.\ (\ref{delrep})
and the remaining symbols are same as in Eq.\ (\ref{ttaver}). 
Adjusting the energy to the isolated electronic level appearing in the
insulating case ($W=11\,a$) at $E=0.04\,t_0$, we immediately find that the
corresponding quantum state is strongly localized in the kink area
(see {\em right} panel in Fig.\ \ref{ldosfig}).
In the metallic case ($W=10\,a$), the value of $E=0.04\,t_0$ belongs
to a~continuum of extended states in the lowest subband, but the corresponding
$\rho_{\rm loc}({\bf R}_j,E)$ profile shows a~clear suppression in the kink area
(see {\em left} panel in Fig.\ \ref{ldosfig}), allowing one to expect that
the current propagation in $y$-direction may be blocked, in the presence
of a~kink,  for a~whole energy
window corresponding to the lowest (or highest) subband for electron 
(or holes).

\section{Conductance and adiabatic pumping in open system}
\label{condpump}

In this section we present central results of the paper concerning
transport properties of the {\em open} system (finite section
of a~nanoribbon attached to the leads) presented in Fig.\ \ref{setup2fig}.

\begin{figure}[!t]
  \includegraphics[width=\linewidth]{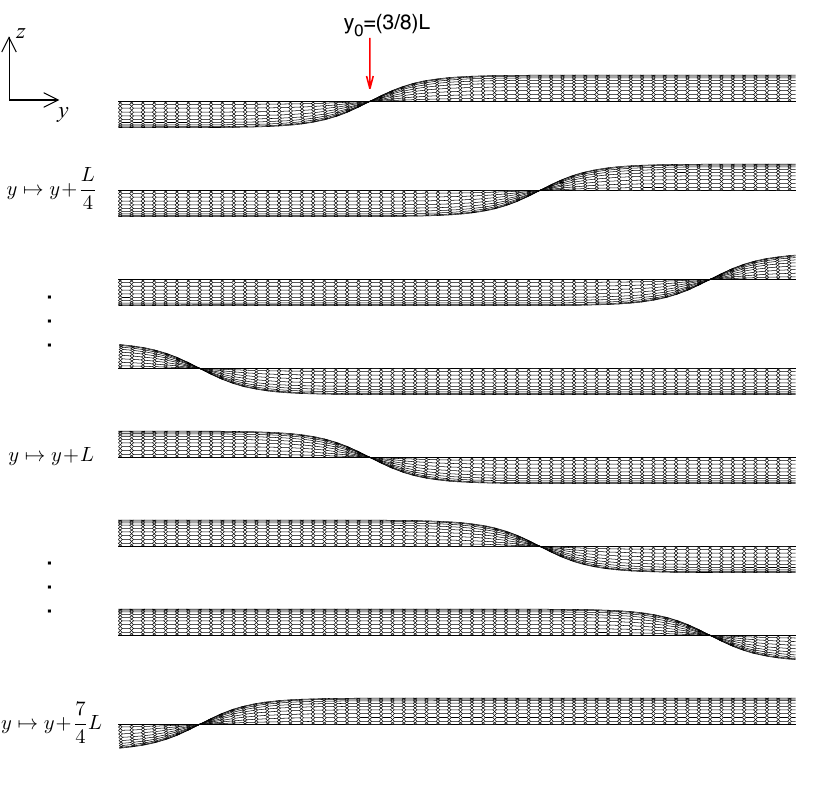}
\caption{ \label{kinkcyclefig}
  (Top to bottom) Buckled graphene ribbon with a~kink, originally placed
  at $y_0=\frac{3}{8}L$, subjected to seven consecutive shifts 
  $y\mapsto{}y+L/4$, visualizing the full pumping cycle ({\em kink} and
  {\em antikink} transition upon $y\mapsto{}y+2L$) as seen in the $(y,z)$
  plane.
  A~modified periodic boundary conditions (i.e., $y\mapsto{}y+L$,
  $z\mapsto{}-z$)
  are applied. The system parameters are $W=11\,a$, $W'/W=0.9$, and
  $L=30\sqrt{3}\,a$. (Notice that a~short ribbon is chosen here for
  illustration only; in the forthcoming calculations we set
  $L=90\sqrt{3}\,a$). 
  The bond lengths are optimized for $\beta=3$. 
}
\end{figure}

\subsection{Simulation details}
So far, we have discussed several characteristics of the {\em closed}
system with modified periodic boundary conditions in the $y$-direction 
(see Sec.\ \ref{qustats}), making the kink position ($y_0$) irrelevant for
global characteristics, such as the density of states. 
Now, we use the atomic positions $\{{\bf R}_j\}=\{(x_j,y_j,z_j)\}$ obtained
with the optimization procedure described in  Sec.\ \ref{model:opt}
(again, we consider the cases without and with the electron-phonon coupling,
$\beta=0$ and $\beta=3$) for $y_0=\frac{3}{8}L$.
Next, the kink is placed at the desired position (say, $y_0+\Delta{}y$)
by applying a~shift to all $y$ coordinates, $y\mapsto{}y+\Delta{}y$.
A~series of consequitive shifts, such as visualized in
Fig.\ \ref{kinkcyclefig}, emulates the kink motion (including full {\em kink}
and {\em antikink} transitions) in a~real system.
In case the shift is commensurate with the longitudinal ribbon periodicity, 
$\Delta{}y=\sqrt{3}na$ with $n$-integer, we simply apply modified
periodic boundary conditions for all atoms, for which $y_j+\Delta{}y<0$ or
$y_j+\Delta{}y\geqslant{}0$.
Otherwise (i.e., if $\Delta{}y\neq{}na$), atomic positions
after a~shift $\{{\bf R}_j\}_{\Delta{}y}$ are determined via 
third-order spline interpolation using $\{{\bf R}_j\}_{\sqrt{3}(n_0-1)a}$,
$\{{\bf R}_j\}_{\sqrt{3}n_0a}$, \dots, $\{{\bf R}_j\}_{\sqrt{3}(n_0+2)a}$,
with $n_0=\lfloor\Delta{}y/(\sqrt{3}a)\rfloor$, and $\lfloor{}x\rfloor$
the foor function of $x$. 

The hopping-matrix elements ($t_{ij}$) in Eq.\ (\ref{ttaver}) are then
determined using atomic positions after a~shift, $\{{\bf R}_j\}_{\Delta{}y}$,
but we set $t_{ij}=0$ in case $i$ and $j$ are terminal atoms from the opposite
zigzag edges (i.e., periodic boundary conditions are no longer applied for
electronic degrees of freedom).

The leads, positioned at the areas of $x<0$ and $x>W'$
in Fig.\ \ref{setup2fig},
are modeled as perfectly flat (i.e., $t_{ij}=-t_0$ for the nearest neighbors
$i$ and $j$) and heavily doped graphene areas, with the electrostatic potential
energy $U_\infty=-0.5\,t_0$ (compared to $U_0=0$ in the ribbon area, $0<x<W'$),
each of the width $W_\infty=17.5\sqrt{3}\,a$ (corresponding to $11$ propagating
modes for $E=0$). What is more, both leads are offset from the free ribbon
edges by a~distance of $L_s=7.5\sqrt{3}\,a$, suppressing the boundary effects. 
The scattering problem is solved numerically, for each
value of the chemical potential $\mu=E-U_0$ and the kink position $y_0$,
using the {\sc Kwant} package \cite{Kwant} in order to determine the
scattering matrix
\begin{equation}
\label{smatrix}
  S(\mu,y_0)=\left(\begin{array}{cc}
    r & t' \\
    t & r' \\
  \end{array}\right), 
\end{equation}
which contains the transmission $t\,$ ($t'$) and reflection $r\,$ ($r'$)
amplitudes for charge carriers incident from left (right)
lead.

\begin{figure}[!t]
  \includegraphics[width=\linewidth]{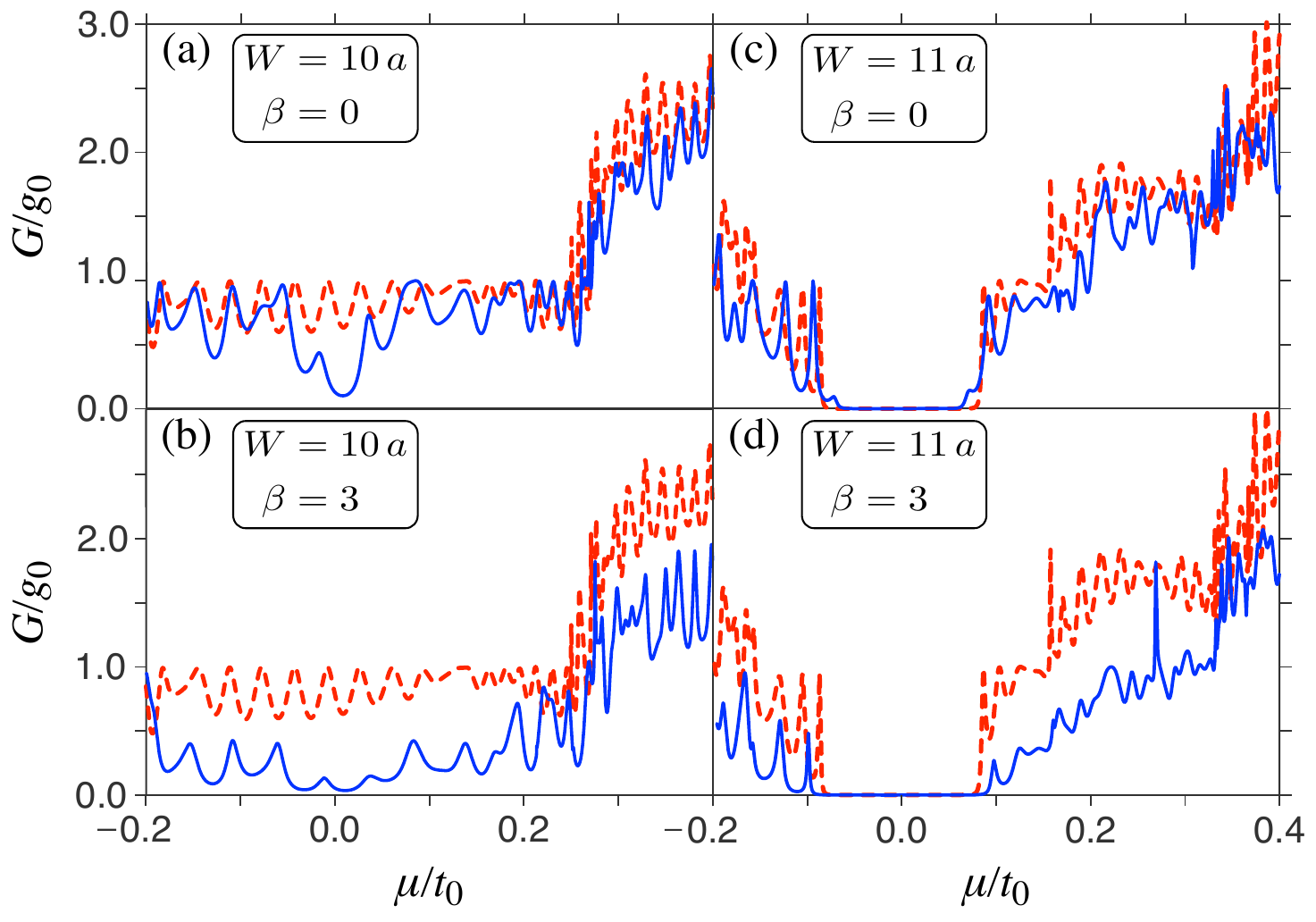}
\caption{ \label{conductafig}
  Conductance of graphene ribbon attached to the leads
  (see Fig.\ \ref{setup2fig}) with $L=90\sqrt{3}\,a$,
  $W_{\infty}=17.5\sqrt{3}\,a$, $L_s=7.5\sqrt{3}a$, and
  $L_1=40\sqrt{3}\,a$, displayed as a~function of the chemical
  potential. 
  The ribbon with is (a,b) $W=10\,a$ or (c,d) $W=11\,a$.
  Blue solid lines in all plots are for $W'/W=0.9$, the kink
  position $y_0=(3/8)\,L$, and the bond lengths
  optimized for (a,c) $\beta=0$, or (b,d) $\beta=3$;
  red dashed lines correspond to a~flat ribbon case ($W'\!=\!W$). 
}
\end{figure}

\subsection{Landauer-B\"{u}ttiker conductance}
The linear-response conductance is determined from the $S$-matrix via
the Landauer-B\"{u}ttiker formula \cite{Lan57,But85}, namely
\begin{equation}
  G=G_0\mbox{Tr}\,tt^\dagger = \frac{2e^2}{h}\sum_nT_n, 
\end{equation}
where $G_0=2e^2/h$ is the conductance quantum and $T_n$
is the transmission probability for the $n$-th normal mode.

In Fig.\ \ref{conductafig}, we compare the conductance spectra for
the same four combinations of parameters $W$ and $\beta$ as earlier used
when discussing the density of states (see Fig.\ \ref{dos4fig}).
This time, results for a~buckled ribbon, with $W'/W=0.9$ and a~kink
placed at $y_0=\frac{3}{8}L$, are compared with the corresponding results
for a~flat ribbon (solid blue and dashed red lines
in Fig.\ \ref{conductafig}, respectively).
In the metallic case, electron-phonon coupling strongly
suppresses the transport in the presence of a~kink
[Fig.\ \ref{conductafig}(b)]; the effect of a~kink is much weaker
in the absence of electron-phonon coupling [Fig.\ \ref{conductafig}(a)].
Similar effects can be noticed in the insulating case, provided
that the chemical potential is adjusted to the first conductance step above
(or below) the gap range [Figs.\ \ref{conductafig}(c) and
\ref{conductafig}(d)].

\begin{figure}[!t]
  \includegraphics[width=0.9\linewidth]{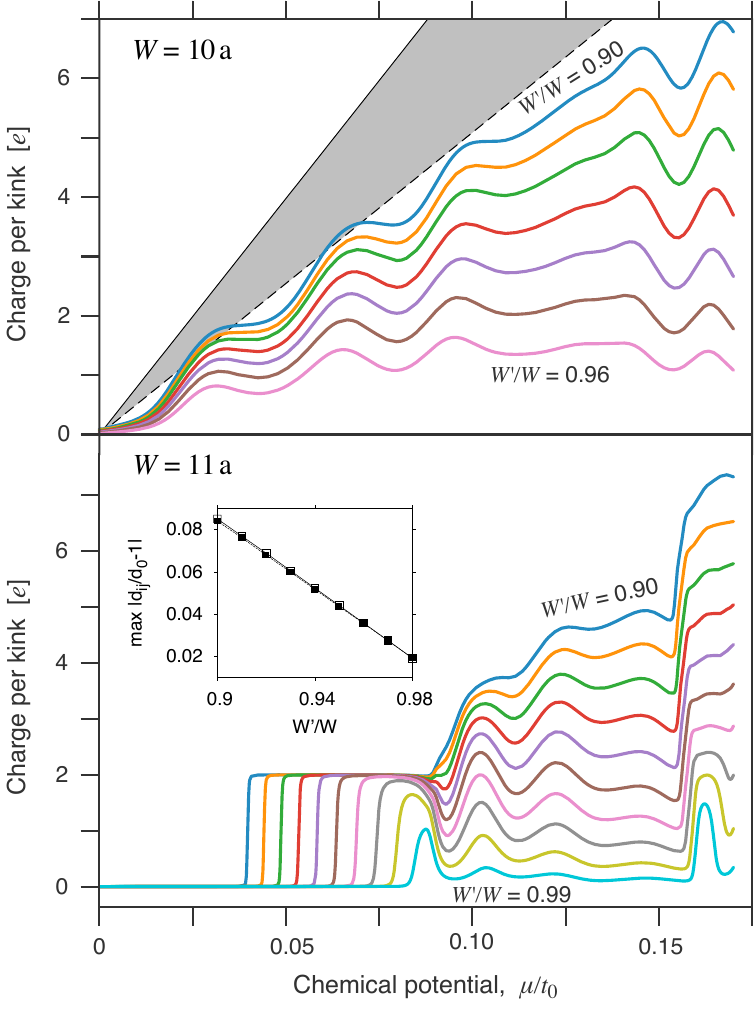}
\caption{ \label{pumpfig}
  Charge pumped per each kink (or antikink) transition 
  as a~function of the chemical potential for  
  (top) $W=10\,a$ and (bottom) $W=11\,a$.
  The bond lengths are optimized for $\beta=3$. 
  The ratio $W'/W$ is 
  varied between the lines with the steps of $0.01$; the maximal
  and minimal $W'/W$ are specified at each plot.
  Shaded area (top) marks the total charge available for pumping, 
  approximated by Eq.\ (\ref{qkinapp})
  with $L_1\leqslant{}L_{\rm eff}\leqslant{}L_1+W_{\infty}$.
  Inset (bottom) shows maximal absolute bond distortion, as a~function 
  of $W'/W$, for $W=10\,a$ (open symbols) and $W=11\,a$ (closed symbols);
  lines in the inset are drawn to guide the eye only. 
}
\end{figure}

\subsection{The pumping spectra}
In the absence of a~voltage bias between the leads, the charge transferred 
solely due to adiabatic kink motion (i.e., by varying the parameter $y_0$)
can be written as \cite{Naz09}
\begin{equation}
  \label{delq}
  \Delta{}Q=-\frac{ie}{2\pi}\sum_{j}\int{}dy_0\left(
  \frac{\partial{}S}{\partial{}y_0}S^\dagger
  \right)_{jj}, 
\end{equation}
where the summation runs over the modes in a~selected ({\em output}) lead.
We further notice that molecular dynamics simulations of
Refs.\ \cite{Yam17,Yam19} allow to estimate typical kink velocity
(up to the order of magnitude) as
$v_{\rm kink}\sim{}1\,$km/s$\ \ll{}v_F$,
where $v_F=\sqrt{3}\,t_0{}a/(2\hbar)\approx{}10^6\,$m/s is the Fermi
velocity in graphene, justifying the adiabatic approximation
\cite{ekinkfoo}. 

Numerical results for $\Delta{}Q(\mu)$, obtained by shifting the kink
from $y_0=0$ to $y_0=L$, are presented in Fig.\ \ref{pumpfig}.
Although the current blocking in the metallic case is far from being perfect
[see Fig.\ \ref{conductafig}(b)], the related pumping mechanism for $W=10\,a$
appears to be rather effective (see top panel in Fig.\ \ref{pumpfig}),
with $\Delta{}Q(\mu)$ approaching the total charge available for transfer
in a~section of the length $L_{\rm eff}$, a~value of which can be approximated
by \cite{qkinkfoo} 
\begin{equation}
\label{qkinapp}
  \frac{Q_{\rm kink}}{e} \approx
  g_s\frac{L_{\rm eff}|\mu_0|}{\pi{}\hbar{}v_F}
  =\frac{4}{\pi}\frac{L_{\rm eff}|\mu_0|}{\sqrt{3}\,t_0{}a},
\end{equation}
where we put $L_1\leqslant{}L_{\rm eff}\leqslant{}
L_1+W_{\infty}$ estimating the effective length of a~ribbon section between
the leads (see shaded area in Fig.\ \ref{pumpfig}). 

Significant changes to the $\Delta{}Q(\mu)$ spectra are observed
in the insulating case of $W=11\,a$ (see bottom panel in Fig.\ \ref{pumpfig}).
Namely, there is an abrupt switching between $\Delta{}Q\approx{}0$ near the
center of a~gap (at $\mu=0$) and $\Delta{}Q\approx{}2e$ appearing for $\mu$
exceeding the energy level localized in the kink area
[see Fig.\ \ref{dos4fig}(d)].
The value of $\Delta{}Q\approx{}2e$ remains unaffected until $\mu$ approaches
a~bottom of the lowest electronic subband [corresponding to the first
conductance step in Fig.\ \ref{conductafig}(d)].
For higher $\mu$, the picture becomes qualitatively similar to this for
a~metalic case, with $\Delta{}Q(\mu)$ systematically growing with $\mu$
and degreasing with $W'/W$. 
Noticeably, the plateau with $\Delta{}Q\approx{}2e$ is well-developed
starting from moderate bucklings, $W'/W\approx{}0.95$.
For $W'/W\approx{}0.9$, deviation from the quantum value in the plateau range
is of the order of $|\Delta{}Q-2e|\sim{}10^{-4}\,e$, and can be attributed
to the finite-size effects.
Some stronger deviations may 
appear in a~more realistic situation due to the finite-temperature
and non-adiabatic effects, which are beyond the scope of this work. 

In both (metallic and insulating) cases, the stability of numerical
integration in Eq.\ (\ref{delq}) substantially improves for the lead offsets
$L_s\gtrsim{}5\,a$ (being comparable with the kink size), for which parts of
the ribbon attached to the leads, together with a~section between the leads,
are (almost) uniformly buckled for either $y_0\approx{}0$ or $y_0\approx{}L$. 

In Fig.\ \ref{pumpfig}, we also display maximal bond distortions for
different bucklings (see the {\em inset}), showing that local deformations
$|\delta{}d_{ij}|<0.1\,d_0$ for all $0.9\leqslant{}W'/W<1$.

\section{Conclusions}
\label{conclu}
We have demonstrated, by means of computer simulations of electron
transport, that buckled graphene nanoribbon with a topological defect
(the {\em kink}) moving along the system may operate as adiabatic
quantum pump.
The pump characteristic depend on whether the ribbon is metallic
or insulating. In the former case, even for moderate bucklings
(with relative bond distortions below $10\%$) the kink strongly
suppresses the current flow, and shifts the electric charge when moving
between the leads attached to the system sides. In turn, the charge
pumped per cycle is not quantized.
For insulating ribbon, there are electronic states localized near 
the kink (with energies lying within the energy gap) which can be utilized
to transport a~quantized charge of $2e$ per kink transition (with the factor
$2$ following from spin degeneracy), providing a~candidate
for the quantum standard ampere. 

Remarkably, the current suppression, and subsequent effects we have
described, are visible after the bond lengths optimization for the
Su-Schrieffer-Heeger model is performed, introducing significantly
stronger bond distortions than the classical (a~molecular-dynamics-like)
model optimization. 
Therefore, electron-phonon coupling appears to be a~crucial factor
for utilizing the moving kink for adiabatic quantum pumping
in buckled graphene ribbons.

\section*{Acknowledgments}
We thank Tomasz Roma\'{n}czukiewicz and Krzysztof Ro\'{s}ciszewski 
for discussions. 
The work was supported by the National Science Centre of Poland (NCN)
via Grant No.\ 2014/14/E/ST3/00256. 
Computations were partly performed using the PL-Grid Infrastructure.



\begin{thebibliography}{}
\bibitem{lamm2}
Interactions between carbon atoms are described using the standard two-body,
three-body, four-body (dihedral), and Lennard-Jones potential energy
terms, see Ref.\ \cite{Yam17}. Periodic boundary conditions in $y$-direction
are not applied in this case.
\bibitem{scalarfoo}
Some scalar fields, modifying the electrostatic potential energy by 
$\delta{}V=g(u_{xx}+u_{yy})$, may 
also appear in the kink area (see Ref.\ \cite{Voz10}) but they are neglected
in our analysis, since the coupling constant is not determined precisely
(various works report $g\approx{}4-20\,$eV); also, the effective
electrostatic potential barrier cannot substantially alter
the current blocking (or the electron trapping) mechanism due to Klein
tunelling.
\bibitem{ekinkfoo}
In particular, local electric-field variation
due to kink transition can roughly be estimated as
$\delta\mbox{\boldmath ${E}$}\sim{}(\hbar/e)\delta\mbox{\boldmath ${A}$}
\times\left(W/v_{\rm kink}\right)^{-1}$, with $W/v_{\rm kink}$ quantifying the
transition time for a~fixed position along the ribbon, 
and $\mbox{\boldmath ${A}$}$ being the vector potential given by Eq.\
(\ref{akstrain}) in the main text.
For the geometry considered, the potential energy variation
is significant only in the $x$ direction and
can be bounded by
$\Delta{}V\lesssim{}eW|\delta{}E_x|_{\rm max} \sim
\hbar{}v_{\rm kink}c\beta\Delta{}u_{xx}/d_0$.
For the strain amplitude $\Delta{}u_{xx}<0.1$ and $W=10\,a$,
one finds the ratio $\Delta{}V/\Delta{}E_W\lesssim{}10^{-3}$,
with the subband splitting given by Eq.\ (\ref{delenw}) in the main text.
\bibitem{qkinkfoo}
Eq.\ (\ref{qkinapp}) follows from the dispersion relation
for lowest subbands in the metallic case (with opposite group velocities),
$E_{\pm}^{(1)}=\pm{}\hbar{}v_F{}k_y$; the contribution from higher subbands
may be suppressed due to mismatching group velocities.
\end{thebibliography}
\end{document}